# Mid-infrared quantum cascade laser frequency combs based on multi-section waveguides

Ruijun Wang*, Philipp Täschler, Filippos Kapsalidis, Mehran Shahmohammadi, Mattias Beck, Jérôme Faist

*Institute for Quantum Electronics, ETH-Zurich, CH-8093 Zurich, Switzerland*
*Corresponding author: ruiwang@phys.ethz.ch*

**We present quantum cascade laser (QCL) frequency comb devices with engineered waveguides for managing the dispersion. The QCL waveguide consists of multiple sections with different waveguide widths. The narrow and wide sections of the waveguide are designed in a way to compensate the group velocity dispersion (GVD) of each other and thereby produce a flat and slightly negative GVD for the QCL. The QCL exhibits continuous comb operation over a large part of the dynamic range of the laser. Strong and narrow-linewidth intermode beatnotes are achieved in more than 300 mA wide operation current range. The comb device features also considerably high output power (>380 mW) and wide optical bandwidth (>55 cm$^{-1}$). © 2020 Optical Society of America**

Optical frequency combs are promising light sources for many applications as they offer an attractive combination of broad wavelength coverage and high spectral resolution [1]. Possessing phase coherent modes, frequency combs provide a direct link between the optical and the radio frequency (RF) domains. Based on these unique features, dual comb spectroscopy arises as a powerful technique that enables broadband and high-resolution spectroscopy combined with a very short acquisition time [2]. Dual comb spectroscopy in the mid-infrared is of great interest in various fields such as environmental and medical monitoring since this spectral range contains strong characteristic vibrational transitions of many important molecules [3-5]. Therefore, the availability of reliable and stable mid-infrared frequency comb sources is very valuable for spectroscopy.

Quantum cascade lasers (QCLs) are semiconductor lasers based on intersubband transitions in quantum wells [6, 7]. Since they can cover the 3-20 µm wavelength range and output Watt-level optical power in continuous-wave (CW) operation at room temperature, QCLs are excellent light sources for spectroscopy in the mid-infrared spectral range [8, 9]. The intersubband transitions produces strong optical nonlinearity, which introduces a broadband four-wave-mixing (FWM) process in the QCL waveguides [10, 11]. The FWM process provides a mechanism enabling the generation of frequency combs in QCLs. In recent years, it has been demonstrated that Fabry–Pérot and ring QCLs can directly produce frequency combs without any extra components [12-17]. In the dynamic range of the lasers, Fabry–Pérot QCL comb devices typically operate in three different regimes: the single-mode regime, the frequency comb regime and the high-phase noise regime [18]. Only in the comb regime, a narrow-linewidth intermode beating can be observed in the RF domain as the outcome of the coherent comb modes beating with each other. In the high-phase noise regime, the beatnotes are much broader than the comb regime. For many Fabry–Pérot QCL devices based on conventional waveguide design, we observe the comb operation regions to be segmented into several small regions separated by the high-phase noise operation ranges, as shown in Fig. S1 in Supplement 1. Besides, the comb operation regime in terms of spectrum and beatnote stability is strongly dependent on the dispersion characteristics of the device [19]. Formation of stable frequency combs in QCLs requires sufficiently large nonlinearity and flat group velocity dispersion (GVD) with low enough value [11, 19]. Recently many efforts have been devoted to dispersion engineering to achieve stable mid-infrared frequency combs in an extended portion of the dynamic range of QCLs [18-23]. One approach is to integrate a Gires–Tournois interferometer (GTI) onto the QCL facet to compensate the dispersion [18, 21]. The GTI usually consists of many layers of dielectric materials (e.g., $Al_2O_3$ and $SiO_2$) and terminates with a gold layer. These coatings are often incompatible with high optical power operation of the devices due to the relatively low thermal conductivity of the deposited dielectric materials. Alternatively, J. Hillbrand *et al.* used a planar mirror placed behind the back facet of the QCL to shape the dispersion [19]. In this approach, the dispersion and the fraction of the comb regime are tuned by the position of the mirror. By properly selecting the mirror position, the dispersion created by the external cavity partially compensates the dispersion of the QCL chip, thereby producing a total GVD with relatively flat curve in the laser operation wavelength range and allowing stable comb operation. Besides, dispersion compensation can be realized by coupling the optical mode to a highly doped cladding layer or an extra passive waveguide [20, 22].

In this letter, we present QCL frequency comb devices based on multiple section waveguides for engineering of the dispersion. The QCL device consists several sections with different waveguide widths. The positive GVD (negative slope) of the narrow waveguides partially offsets the negative GVD (positive slope) of the narrow waveguide and thereby introducing a flat and negative GVD for the QCL device.

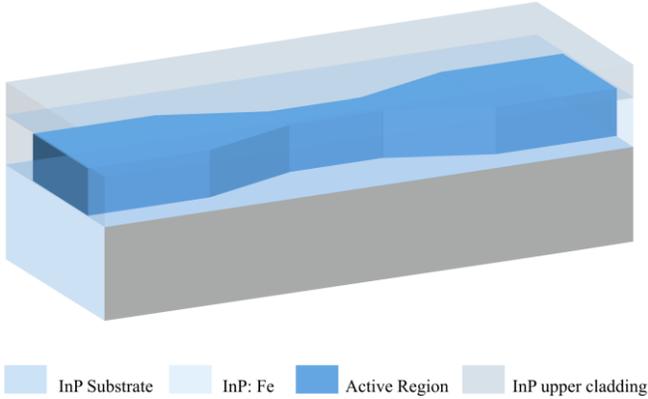

Fig. 1. Schematic drawing of the QCL with a multiple section waveguide.

The schematic drawing of the multiple section QCL waveguide is shown in Fig.1. The waveguide has five sections, including two wide waveguide sections on both sides, one narrow waveguide section at the center and two tapered sections used to connect the wide and narrow waveguide sections. The wide waveguide is capable of supporting the fundamental transverse mode and some higher-order transverse modes while the central narrow waveguide only can support the fundamental mode. The wide waveguides are gradually tapered to a narrow waveguide to minimize radiation loss in the conversion of high-order modes to the fundamental mode. From an optical simulation using finite difference eigenmode method, the transmission efficiency at 8 um wavelength is more than 95% when the QCL waveguide width increases from 3 µm to 20 µm in a 500 µm-long tapered waveguide. In such multiple section waveguides, the higher-order modes are filtered by the central single mode waveguide and the narrow part of the tapered waveguides. Meanwhile, the central narrow section increases the optical field intensity thereby enhancing the optical nonlinearity. Similar structures have been used in fibers to control the nonlinearity and dispersion. For example, tapering sulfide-based chalcogenide fiber leads to a zero-dispersion and enables the generation of an octave spanning supercontinuum spectrum using ultralow pump pulse energy [24]. For the QCL presented in this work, the width of the wide waveguides and narrow waveguide is 12 µm and 4 µm, respectively. The total length of the wide waveguide sections is 2.5 mm while the central narrow waveguide and tapered waveguide sections are all 0.5 mm long.

Figure 2(a) shows the simulated GVD of the fundamental mode of two straight QCL waveguides with different widths. Material dispersion and waveguide dispersion are included in the calculation. It can be seen that a narrower waveguide exhibits a more positive GVD. When the waveguide is 4 µm wide, its GVD value is higher than 900 fs$^2$/mm in the 1200-1300 cm$^{-1}$ wavelength range. As the waveguide width increases to 12 µm, the GVD turns to negative in the same spectral range. This distinct difference in GVD can be ascribed to the dependence of modal dispersion on the waveguide width. For a very narrow waveguide, e.g., a 4 µm-wide waveguide for an 8 µm wavelength QCL, the modal confinement is highly wavelength-dependent in the laser operation wavelength range. Therefore, the GVD is dominated by modal dispersion of the waveguide. When the waveguide width increases, the modal confinement become stronger, and the wavelength dependence of the modal confinement reduces. Therefore, the GVD is instead dominated by material dispersion. For the 8µm wavelength range, the material dispersion of InP and related III-V materials is negative [21]. As a result, the GVD of the QCL waveguide reduces as the width increases.

To verify the effect of multiple section waveguide on dispersion, the GVD of the waveguide-engineered QCL was measured and compared with a QCL based on an 8 µm-wide straight waveguide. For straight-waveguide QCLs operating in the 8 µm wavelength range, 6-8 µm wide is the optimal waveguide width for comb operation [25]. Narrower waveguides introduce higher GVD and less fraction of comb regimes while wider waveguides give rise to lateral multimode lasing. The GVD measurements were carried out when the devices were driven a few tens of miliamperes lower than the threshold current. The GVD value is extracted by analyzing the interferogram of subthreshold operation by a Fourier–Transform approach [26]. The deduced GVD of the QCL with a multiple section waveguide and straight waveguide are shown in Fig. 2(b). The straight waveguide device is operating with a negative GVD between -450 fs$^2$/mm to -750 fs$^2$/mm in the laser operation spectral range with a strong wavelength dependence, which was shown experimentally to be unfavorable for the generation of stable comb in a large portion of the dynamic range of the QCL [18]. By implementing the multiple section waveguide design, the negative slope GVD of the narrow waveguide section offsets positive slope GVD of the wide waveguide section thereby producing a flat GVD for the whole waveguide, as shown in the inset of Fig.2 (a). As a result, the QCL with multiple section waveguide exhibits a flat GVD with a total negative value in the range between -450 fs$^2$/mm to -400 fs$^2$/mm in the laser operation wavelength range as shown in Fig.2(b).

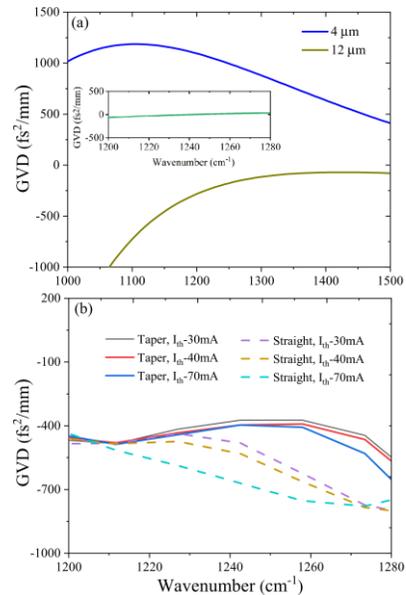

Fig. 2. (a) Simulated GVD of the fundamental mode of a 4 µm and 12 µm-wide straight waveguide. The inset shows the simulated average GVD value of the multiple section waveguide in the QCL lasing spectral range. (b) The comparison of the measured GVD of a QCL with multiple section waveguide (solid lines) and straight waveguide (dashed lines). Both devices were characterized at currents of 30 mA, 40 mA and 70 mA lower than the threshold current.

The active region of the QCL presented in this work is based on a strain-compensated InGaAs/AlInAs heterostructure. The epitaxial layer was grown by molecular beam epitaxial on an InP wafer (Si doped, n=1.3×10$^{17}$ cm$^{-3}$). The active region is sandwiched between two 400 nm-thick InGaAs waveguide layers. The wafer was processed into QCL ridge waveguides using a standard buried heterostructure technique [7]. A layer of InP:Fe was grown on the sides of ridge waveguide for the electrical insulation. After ridge waveguide fabrication, a 2.0 µm-thick InP layer (Si doped, n= 1×10$^{16}$ cm$^{-3}$), a 0.5 µm-thick InP layer (Si doped, n=5×10$^{16}$cm$^{-3}$), a 0.5 µm-thick InP layer (Si doped, n= 5×10$^{17}$ cm$^{-3}$) and a 0.5 µm-thick InP layer (Si doped, n= 1×10$^{19}$ cm$^{-3}$) were grown by metalorganic-vapor-phase epitaxy on top of the upper InGaAs layer to serve as upper waveguide cladding [20]. The QCL devices were mounted epilayer-side down on an AlN and copper heat sink. Both facets of the presented QCL devices are uncoated.

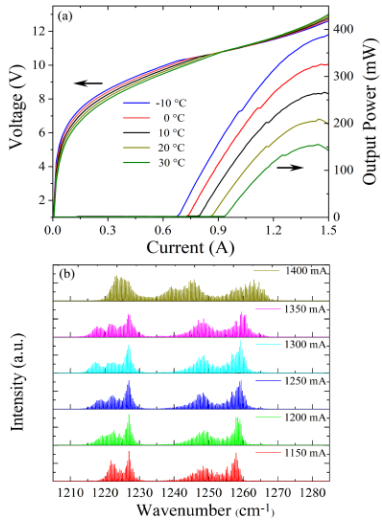

Fig. 3. (a) Light-current-voltage characterization of the studied multiple section QCL in CW operation at heat-sink temperatures from -10 °C to 30 °C. (b) Lasing spectra of the QCL at different driving current at -10 °C.

The QCL based on the multiple section waveguide was characterized in CW and pulsed operation. The device was mounted in a thermoelectric cooler box during the measurements. The light is collected from a single facet. Figure 3(a) shows the measured light-current-voltage (L-I-V) characterization of the 4 mm-long Fabry–Pérot QCL in a CW regime from -10 °C to 30 °C. This device outputs a CW power up to 385 mW at -10 °C with a threshold current of 680 mA, which corresponds to a current density of 1.7 kA/cm$^2$. When the heat-sink temperature increases to 20 °C, the maximum output power decreases to 205 mW and the threshold current density increases to 2.1 kA/cm$^2$. In a pulsed regime (pulse duration 104 ns, period of 10.4 µs), the QCL has a threshold current density of 1.55 kA/cm$^2$ and a maximum (peak) output power of 930 mW at 20 °C. By comparing the threshold current in CW and pulsed operation, a thermal resistance of 6.1 K/W is extracted [7]. Lasing spectra of the devices in CW operation were acquired by Fourier-transform infrared spectrometer with a spectral resolution of 0.075 cm$^{-1}$ at different current levels at -10 °C, as shown in Fig. 3(b). At bias current of 1250 mA, the spectra span a wavelength range of more than 55 cm$^{-1}$.

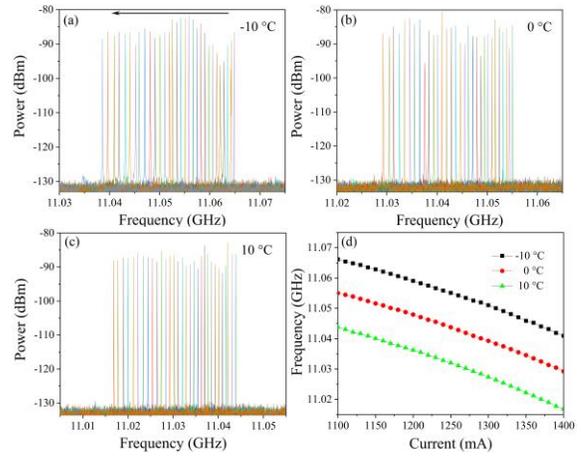

Fig. 4. (a)-(c) Electrically detected intermode beatnote spectra of the QCL when the driving current increases from 1.1 A to 1.4 A with a current step of 10 mA at (a) -10 °C, (b) -0 °C and (c) 10 °C. A spectrum span of 100 MHz and resolution bandwidth of 100 Hz was used in the measurement. The arrow in Fig. 4(a) indicates the direction of the driving current increases. (d) The extracted beatnote frequency from part (a-c) as a function of the driving current. Note the continuous tuning of the intermode beatnote.

In order to evaluate the frequency comb performance of the QCL based on the multiple section waveguide, the intermode beat notes were measured using an electrical spectrum analyzer (ESA). In the measurement, the current driver, QCL and ESA were connected to the DC, DC+RF and RF port of a bias-T, respectively. At -10 °C, a narrow-linewidth beatnote appears when the driving current increases to around 1.08 A, which overlaps the position of the mild kink in the L-I curve. Figure 4(a)-4(c) shows the measured beatnote spectra for different driven currents when the device was operating at a heat-sink temperature from -10 °C to 10 °C. The spectra were firstly acquired with a span of 100 MHz and resolution bandwidth (RBW) of 100 Hz. Very sharp and strong (>35 dBm) beanote spectra can be observed in the whole current range of 1100–1400 mA at different temperatures. The beatnote frequency is linearly tuned by the driven current with tuning coefficient of 85 kHz/mA at -10 °C and 90 kHz/mA at 10 °C as shown in Fig. 4(d). To accurately measure the linewidth, the beanotes at different currents were acquired by ESA with a span of 200 kHz and RBW of 50-300 Hz. Figure 5(a) shows the beatnote spectrum with a span of 100 kHz and a RBW of 50 Hz, for a current of 1250 mA at -10 °C. The beanote exhibits a very narrow linewidth around 100 Hz. In the current range of 1100–1400 mA, all beatnotes have a full width at half maximum less than 1.5 kHz as shown in Fig. 5(b). In contrast, two QCLs with a 6 µm-wide and 8 µm-wide straight waveguide (processed in the same wafer) show narrow-linewidth beatnotes in much narrower operation current range (<100 mA current span) as shown in Fig. S1. These results indicate that QCLs having stable frequency comb generation with improved operation regime are achieved by employing the multiple section waveguide design.

The performance of another five QCL devices based on the multiple section waveguides are shown in the Supplement 1. All devices exhibit narrow beatnotes in a very wide operation current ranges. For example, as shown in Fig. S2, a 4 mm-long QCL with two 17µm-wide waveguide sections can generate narrow-linewidth intermode beatnotes in >500 mA operation current span.

Figure S7 shows the image of the output beam of a multiple section QCL. The optical beam only consists of a zero order mode. This result indicates the multiple section waveguide design can prevent the lasing of higher-order transverse modes.

Fig.5. (a) Beatnote spectrum with a span of 100 kHz and RBW of 50 Hz, for a current of 1250 mA at -10 °C. (b) Beatnote linewidth as a function of the driving current at -10 °C.

To further access the coherence properties of QCLs based on the multiple section waveguides, we performed a correlation analysis by means of coherent beatnote spectroscopy [13, 27−29]. The correlation magnitude of the electric field E(ω) and its frequency shifted replica E(ω+Δω) were obtained by two different measurement methods, once using conventional Fourier transformer spectroscopy (FTS), once using shifted wave interference Fourier transform spectroscopy (SWIFTS). Detailed information about SWIFTS technique can be found in [27]. Figure 6 displays a comparison of the correlation magnitudes obtained by these two methods when the device was driving in the same condition. A good overall agreement can be seen in a spectral coverage around 55 cm$^{-1}$, despite the fundamentally different nature of the measurements. These results indicate that essentially all of spectral power contributes to the frequency comb operation.

Fig.6. Spectrum product obtained by FTS overlaid with the SWIFTS spectrum, demonstrating the excellent coherence of this comb.

In conclusion, we have demonstrated QCL frequency comb devices based on the multiple section waveguide design. A single-facet output power of 380 mW and wavelength coverage more than 55 cm$^{-1}$ have been achieved. The multiple section waveguide geometry can be employed to engineer the dispersion of QCL frequency combs. The positive dispersion (negative slope) of the narrow waveguide compensates the negative dispersion (positive slope) of the wide waveguides and thereby leading to a flat and negative GVD. The QCLs with multiple section waveguides output stable frequency combs in a large and continuous portion of the dynamic range of the lasers. The frequency comb operation is verified by coherent beatnote spectroscopy measurements.

**Funding**. Schweizerischer Nationalfonds zur Förderung der Wissenschaftlichen Forschung 176584.

**Acknowledgments**. The authors would like to thank Dr. Emilio Gini for the MOVPE regrowth, Matthew Singleton and Zhixin Wang for his careful reading this paper and helpful suggestions.

**Disclosures**. The authors declare no conflicts of interest.

See Supplement 1 for supporting content.